\documentclass[aps,twocolumn]{revtex4-1}

\usepackage{bm, amsmath,amssymb,amsfonts,slashed,array,graphicx,soul}
\usepackage{mathrsfs}
\usepackage{xcolor}
\usepackage{hyperref}
\usepackage{cancel}
\usepackage[normalem]{ulem}
\usepackage{amsmath}
\usepackage{amsfonts}
\usepackage{amssymb}
\usepackage{array,booktabs}
\usepackage{float}
\usepackage{slashed}            
\usepackage{tikz}
\usetikzlibrary{arrows,shapes}
\usetikzlibrary{trees}
\usetikzlibrary{matrix,arrows}              
\usetikzlibrary{positioning}                
\usetikzlibrary{calc,through}               
\usetikzlibrary{decorations.pathreplacing}  

\usetikzlibrary{decorations.pathmorphing}   
\usetikzlibrary{decorations.markings}
\usetikzlibrary{snakes}
\usepackage[normalem]{ulem}

\newcommand{\nc}{\newcommand}

\nc{\beq}{\begin{equation}} \nc{\eeq}{\end{equation}}
\nc{\beqa}{\begin{eqnarray}} \nc{\eeqa}{\end{eqnarray}}
\nc{\bea}{\begin{eqnarray}} \nc{\eea}{\end{eqnarray}}
\nc{\barray}{\begin{eqnarray}} \nc{\earray}{\end{eqnarray}}
\nc{\barrayn}{\begin{eqnarray*}} \nc{\earrayn}{\end{eqnarray*}}
\nc{\ra}{\rightarrow}
\nc{\lsim}{\begin{array}{c}\,\sim\vspace{-21pt}\\< \end{array}}
\nc{\gsim}{\begin{array}{c}\sim\vspace{-21pt}\\> \end{array}}
\nc{\Tr}{{\rm Tr}} \nc{\slsh}{\slash\hspace*{-0.22cm}}

\def\be{\begin{equation}}
\def\ee{\end{equation}}
\def\bea{\begin{eqnarray}}
\def\eea{\end{eqnarray}}
\def\bit{\begin{itemize}}
\def\eit{\end{itemize}}

\nc{\infinity}{\infty} \nc{\mc}{\mathcal} \nc{\M}{\mathcal{M}}

\def\lsim{\mathrel{\rlap{\lower4pt\hbox{\hskip1pt$\sim$}}
    \raise1pt\hbox{$<$}}}
\def\gsim{\mathrel{\rlap{\lower4pt\hbox{\hskip1pt$\sim$}}
    \raise1pt\hbox{$>$}}}

\def\to{\rightarrow}

\begin{document}

\title{Faking SM Photons by Displaced Dark Photon Decays}
\author{Yuhsin Tsai$^{1}$, Lian-Tao Wang$^{2,3}$, and Yue Zhao$^{4}$}
\affiliation{$^1$Maryland Center for Fundamental Physics, Department
of Physics, University of Maryland, College Park, MD
20742\\$^2$Department of Physics, The University of Chicago,
Chicago, IL 60637\\$^3$Enrico Fermi Institute and Kavli Institute
for Cosmological Physics, The University of Chicago, Chicago, IL
60637\\$^4$Michigan Center for Theoretical Physics, University of
Michigan, Ann Arbor, MI 48109}
\date{\today}
\begin{abstract}
A light meta-stable dark photon decaying into collimated
electron/positron pair can fake a photon, either converted or
unconverted, at the LHC. The detailed object identification relies
on the specifics of the detector and strategies for the
reconstruction. We study the fake rate based on the ATLAS(CMS)
detector geometry and show that it can be O(1) with a generic choice
of parameters. Especially, the probability of being registered as a
photon is angular dependent. Such detector effects can induce bias
to measurements on certain properties of new physics. In this
paper, we consider the scenario where dark photons in final states
are from a heavy resonance decay. Consequently, the detector effects
can dramatically affect the results when determine the spin of a
resonance. Further, if the decay products from the heavy resonance
are one photon and one dark photon, which has a large probability to
fake a diphoton event, the resonance is allowed to be a vector. Due
to the difference in detector, the cross sections measured in ATLAS
and CMS do not necessarily match. Furthermore, if the diphoton
signal is given by the dark photons, the SM $Z\gamma$ and $ZZ$ final
states do not necessarily come with the $\gamma\gamma$ channel,
which is a unique signature in our scenario. The issue studied here
is relevant also for any future new physics searches with photon(s)
in the final state. We discuss possible ways of distinguishing dark
photon decay and real photon in the future.
\end{abstract}
\maketitle

\section{Introduction}

The large hadron collider (LHC) is currently the most energetic high
energy facility running in the world. The center of mass energy of
the proton-proton collision has reached 13 TeV, and it provides us
direct probes to new physics beyond standard model on energy
frontier. To extract information from each collision, final states
are registered on detectors and further go through highly
non-trivial object reconstruction procedures in order to be
identified. Thus the conclusions one draws from the LHC highly
depend on the object reconstruction strategies. It is interesting
and important to investigate the possible loop holes in object
identification procedures which may lead us to a conclusion
different from what really happened.

Photon is one of the most important objects from collider physics
point of view, especially on a proton-proton collider where dominant
final states are QCD hadrons. In this paper, we focus on the
possibilities on faking the photon signature by exotic hidden sector
decays. The first possibility is to have two or more collimated
photons in final states. Such collimated photons can be the decay
products of a highly boosted light particle. A close analogy happens
in Standard Model (SM) is the highly boosted pion decaying to two
photons through anomaly vertex. Such channel has been checked
carefully in content of Higgs decay. Instead of pion in SM, a light
particle in hidden sector, which is easily isolated with other
particles in final states, can also decay to collimated photons.
This possibility has been discussed in several places (e.g.,
\cite{Dobrescu:2000jt,Chang:2006bw,Larios:2001ma,Larios:2002ha,Toro:2012sv,Draper:2012xt,Ellis:2012sd,Ellis:2012zp,Curtin:2013fra,Chang:2015sdy,Aparicio:2016iwr}). It is also shown in several works
\cite{Knapen:2015dap,Agrawal:2015dbf} that making a highly boosted
(so the two collimated photons are too close to be resolved) hidden
scalar decay into two SM photons within the detector is challenging,
and the charged particles that are introduced to decay the hidden
scalar into SM photons can be highly constrained.

Another way of faking the photon signal, which we explore in this
work, is through a displaced decay into charge leptons. If a light
hidden sector particle decays to a pair of collimated electron and
positron at a distance comparable to the detector size, such objects
have a good chance to fail the lepton tagging and be identified as a
photon at the LHC. Particularly, when a real photon interacts with
materials and converts into an electron/positron pair, the photon
gives the same final state particles as from the hidden sector
decay. More interestingly, as we will show, due to the subtle object
identification strategies in detector, the angular distribution of
the fake photon signal depends on the dark photon lifetime, i.e. the
detector geometry can introduce a bias to signal angular
distribution. Such a bias can be important in determining the
property of new physics. For example, if one or more dark photons
are produced due to a heavy particle decay, studying the angular
distribution of the signal events is the most common way to
determine the spin of the resonance. However, the non-trivial angular distribution of
the fake photon events can change the result in such an analysis and give a different result of the resonance spin.

In this paper, we focus on the scenario where displaced dark
photon decays as $\gamma'\to e^+e^-$. Some preliminary ideas about
the photon faking from the dark photon decay is discussed in \cite{Agrawal:2015dbf,Dasgupta:2016wxw}, which focused on
the possibility that the decay of dark photon only fakes a converted
photon. In \cite{Dasgupta:2016wxw}, it is suggested that if a photon signal really comes from a dark
photon decay, one can distinguish it from the SM photon by checking
whether the signal is more populated in the converted photon category.

However, we show in this paper that the above conclusion may not be complete. Through a
more careful study of the photon identification criteria and detector geometry, we
show the dark photon decay can also fake unconverted photon signals. Specifically, for
some generic choices of parameters, the relative rates for dark photons to be
tagged as converted versus unconverted photons can be
similar to SM photons. Thus by simply checking the ratio of
converted/unconverted photons in the signal region may not be
enough to distinguish the two final states \footnote{Note, the fact that
both converted and unconverted photon can be faked by a meta-stable
light particle decaying to $e+/e-$ was also pointed out in
\cite{Bi:2016gca,Chen:2016sck}. However, we take a different
choice of borderline to divide converted photon v.s. unconverted
photon. The difference comes from whether stand-alone TRT tracks can
be used to distinguish converted and unconverted photons. It is
important to notice that reconstruction efficiency of track by TRT
decreases when increasing $p_T$, as shown in Fig.~6 on p.~119 of \cite{Aad:2009wy}. Since we are mainly focused on the high energy regime of
the diphoton resonance search, we expect TRT track reconstruction
efficiency is very low and thus negligible. This is one of the
crucial differences in our study.}.

The outline of this paper is as follows. In Sec.~\ref{sec:detector}, we discuss the details on photon
identification and show how a dark photon decaying at different part
of the detector can fake a photon, either converted or unconverted.
In Sec.~\ref{sec:identification}, we classify the dark photon decay
identification into four categories and study the probability of
each category for generic choices of parameters. We estimate the
ratio between the converted and unconverted photon events, which can
be used to exam the fake photon signal. Also we compare the
kinematics of electron/positron from a dark photon decay to that
from a photon conversion and show they can be similar to each other.
In Sec.~\ref{sec:models}, we consider a benchmark model where a
heavy resonance decays to two dark photons or one SM photon plus one
dark photon. We discuss several interesting model building subtleties
in these cases. In Sec.~\ref{sec:simulation}, we carry out a detailed study on dark photon
parameter space and show that a generic choice of
dark photon mass and mixing can easily give a signal
rate in a diphoton search to be comparable to the SM background. Thus the fake photon signal is important to be taken into
consideration in certain photon-related measurements. We also present the
distortion of angular distributions after taking into account of the
detector effects. Several new physics searches, such as lepton jet
searches and mono-photon search, may be applied to constrain our
parameter space. We find our scenario can be easily consistent with
these searches. At last in Sec. \ref{sec:discussion}, we summarize
and discuss the possible ways to distinguish our scenario from the
conventional ones.

\section{Photon selection criteria}\label{sec:detector}
When a highly-boosted
and meta-stable particle decays to a collimated $e^+e^-$ in the
tracker, there is a good chance for the object to be identified as a
photon. In order to estimate the probability of registering the
meta-stable light particle decay as a photon, we first discuss the
criteria on the photon identification at both ATLAS and CMS.

We note that there can be several differences between the fake and
real converted photons. First, the two leptons from a dark photon
decay can have a different energy distribution to the one from a
photon conversion. Moreover, the probability of a dark
photon being registered as converted/unconverted photon depends on
its lifetime, which can be different from the conversion rate of SM
photons. The detailed location of photon conversion should also
follow the distribution of the material, while the dark photon decay
is not bonded to this. To observe the differences, we need either a
large number of signal events, or a more careful study of the converted $e^+e^-$
kinematics as we will discuss in the later sections.

Whether the dark photon decay is identified as a photon, and whether
it fakes a converted or unconverted photon all depend on the decay
location and the angle between the electron/positron in the final
state. Taking these into account, we investigate the possibility of
obtaining a sizable photon faking probability that satisfies
existing bounds on the various displaced signals. Our discussion
will be mainly based on the details of ATLAS detector
\cite{Aad:2008zzm,ATLAS-CONF-2012-123}. The tracker at the CMS has a
similar geometry to that at the ATLAS, but the angular resolution of
the Electromagnetic Calorimeter (ECAL) in the $(\eta,\,\phi)$
direction is worse in distinguishing the columnated $e^+e^-$ from a
single photon. Therefore, the probability for
a dark photon to fake a SM photon at the CMS is
likely to be higher than that at the ATLAS. We discuss more details
of the CMS detector later in this section.

A converted photon is a cluster of energy deposition in the ECAL
associated with one or two tracks which appear in the middle of
tracker. More precisely, the reconstructed track should start after
the 1st layer of the tracker, which is about 34 mm away from the
central axis of the detector in the barrel region. The number of
tracks can be either 1 or 2 depending on whether the photon
conversion is symmetric or asymmetric. Furthermore, in order to seed
the track, there needs to be at least 3 space points in either Pixel
or Semiconductor Tracker (SCT) \footnote{This is only a requirement
on seeding a track candidate. A more stringent and complicated
analysis will be carried in order to reconstruct a track. In the
following discussion, we will assume all tracks can be reconstructed
once they are seeded. This will introduce a bias which enlarges the
probability of identifying our meta-stable particle decay as a
converted photon instead of a normal photon, as discussed in later
context.}. Thus any electron/positron appears after the
3rd-to-the-last layer of the SCT (it is about 371 mm away from the
central axis in the barrel region) will not leave a reconstructible
track. We emphasize that a converted photon cannot have a track
which registers a hit at the first layer of Pixel in the tracker.
This is because the conversion can only happen in places with
materials. Thus for a meta-stable particle, if its decay happens
between the 1st layer of Pixel and the 3rd-to-the-last layer of SCT,
it will be identified as a converted photon. The converted photon
can in principle be constructed in the Transition Radiation Tracker
(TRT), but the reconstruction efficiency is expected to be much
lower for high energy photons. Different from the clear tracks
obtained in Pixel and SCT, which allow the photon reconstruction
using a single electron (asymmetric conversion), the standard TRT
tracks are much less reliable and need both of the tracks to
identify a converted photon. If the photon $p_T$ is too large, the
electron/positron are hardly separated in the TRT. This is why the
$20$ GeV converted photon has a much lower efficiency than the one
of the $5$ GeV photon, as is shown in Fig.~6 of \cite{Aad:2009wy}.
Hard $p_T$ cuts are usually
imposed in a typical diphoton resonance search. For instance, the leading
photon needs to have $p_T\gsim100$ GeV in \cite{ATLAS-CONF-2016-059, Khachatryan:2016hje},
and it is very unlikely that a converted
photon can be constructed by hits at the TRT.

It is a little bit subtle if the meta-stable particle decays after
the 3rd-to-the-last layer of the SCT. If the decay happens before
the 1st layer of the ECAL, which is about 1500 mm away, there are no
tracks associated to the energy deposition in the ECAL.
What kind of object will the decay signal be
identified as depends on the separation between electron/position
when they hit the ECAL.

The first layer of the ECAL in ATLAS has extremely good resolution on
pseudo-rapidity $\eta$, $\Delta\eta\sim O(10^{-3})$. If the $\eta$
separation of energy deposition from electron/position cannot be
resolved, such an event will be identified as a normal photon. If the
separation can be resolved, the signal structure is similar to the one
from a neutral pion decaying into 2 photons. Such an object will not be
identified as a single photon. If the separation is very large, so that the
energy deposition is grouped as 2 clusters in the ECAL, the final state is
identified as 2 photons close to each other. The last possibility is
very unlikely to happen in our scenario since we assume the
meta-stable particle is very light and thus highly boosted, so we will not consider this
possibility in the later discussion.

The precise angular cut at the first layer of the ECAL, which determines if the signal
is a single photon or not is complicated and $\eta$-dependent. In order to proceed,
we make a rather conservative assumption to simplify the calculation by requiring the separation
$\Delta x$ in length in the first layer of the ECAL to be within the
average width of the strip cells ($\Delta x<4.7$ mm). The bound is
similar to the one discussed in \cite{Knapen:2015dap}, which uses the
argument that a $\pi^0$ with $65$ GeV energy can be distinguished
from a photon in Higgs searches. There is one subtle difference,
however, between the two assumptions. The $\pi^0$ decays promptly in
the detector, while our meta-stable particle needs to decay after
3rd-to-the-last layer of SCT. The spatial separation from the hidden
particle decay is smaller than that of the $\pi^0$ decay if
they share the same opening angle.

In the CMS detector, angular resolution of the lead tungstate
crystals in the ECAL is $\Delta\eta\sim O(10^{-2})$, and a dark photon
decay that passes the angular cut at ATLAS can also look like a
single photon in the CMS. To identify the converted photon, the
Converted Safe Electron Veto (CSEV) used in the CMS also removes
tracks with a hit in the innermost tracker layers
\cite{Gonzi:2014hbs}. The first layer is about $44$ mm away from the
beam pipe, which is similar to the $34$ mm in the ATLAS setup for
our purpose. The crystal calorimeter takes signals up to $1.79$ m,
which is not too far from the requirement at ATLAS of having the
ECAL signal before $1.5$ m. Thus the faking rate for a meta-stable
dark photon decay can be comparable to each other in two
experiments, with only mild differences.

At last, let us discuss more details about the possibility that dark
photon decays either before the tracker or after the 1st layer of
the ECAL. If the decay happens before the tracker, the object will
not be identified as a converted photon since the reconstructed
tracks have hits on the 1st layer of Pixel. In this case, search of
the prompt lepton jets \cite{Aad:2015sms} can be useful in
constraining the scenario, as we will study in
Sec.~\ref{sec:simulation}. If the decay happens after the 1st layer
of the ECAL ($1590<r$ mm in ATLAS), there is a large chance that
such decay will not be constructed as a well defined object
\footnote{This particular definition of MET is widely used in
many other searches involving missing energy. See also \cite{CMS:2009nxa} for the MET reconstraction at CMS. The benefit for doing
this is because that, for a well-reconstructed object, its
transverse momentum can be properly calibrated according its
identity. However, if an object is not well defined in a
particle-flow event reconstruction, such as a dark photon decaying
after ECAL. This object is unlikely to be included when calculating
MET.}. In this case, since the search of monophoton in ATLAS
\cite{Aad:2014tda} defines the MET to be the imbalance of the total
momentum of the well constructed objects, the resonance decay into
two dark photons -- with one has later decay after the ECAL and one
inside the tracker -- may be considered as a monophoton signal. We
study the bound in Sec.~\ref{sec:simulation} under this conservative
assumption. Further, there are specially designed searches if the
decay happens inside the Hadron Calorimeter (HCAL, $>1970$ mm in
ATLAS) for example displaced lepton signals at the HCAL can be
applied \cite{ATLAS-CONF-2014-041,Aad:2014yea}. However, due to
strong requirements on event selection, especially requiring both
meta-stable particles decay in HCAL, these searches only constrain
$8$ TeV dark photon pair production at pb level. This is a much
weaker constraint comparing to the monophoton search.

\section{Object identification summary}\label{sec:identification}
Here we summarize the object identification criteria used in this work. As we
discussed above, whether the dark photon decay is identified as
a photon, especially whether it is a converted or unconverted photon
both highly rely on the details of the detector. A precise simulation
on the detector response is beyond the scope of this paper. Here we carry
out a simplified procedure to estimate how the lifetime of
meta-stable particles affect the object identification.

We classify the meta-stable particle decay into
four regions:

{\bf{Before Tracker:}} If the decay happens before the first layer of the tracker,
i.e. with a distance $\lsim34$ mm in ATLAS, the electron/positron can be identified,
and the signal is not identified as a photon. The non-isolated leptons from a highly
boosted light particle decay (with a boost factor $\sim\mathcal{O}(10^3)$) can be
studied in the lepton-jet search, and we will show in Sec.~\ref{sec:simulation} how
the existing search constrains the parameter space.

{\bf{Converted $\gamma$ Region:}}  If the decay happens within tracker and can leave
at least three space points in either pixel or SCT, we identify such
decay as a converted photon. The details on ATLAS detector is taken
from \cite{website}. We emphasize that this is
an oversimplified criteria for identifying converted photons, and having at least
three space points in Pixel or the SCT is only a minimal requirement to
seed a track. In a more complete search, there can be additional procedures on the track reconstruction.

{\bf{Unconverted $\gamma$ Region:}}  If the decay happens before the ECAL and after the tracker region
so that there are no more than 3 layer of the SCT to be available to reconstruct the
track, we identify the decay signal to be an unconverted photon. This is
under the assumption that the open angle between electron/position from the
meta-stable particle decay is too small to be resolved
as 2 photons in the ECAL. We will see in the later discussion that the angular cut can
be satisfied in the model we are interested in.

{\bf{ECAL $\&$ after:}}  If the decay happens within or after the
ECAL, we do not identify such an object as a photon. What it will be
identified depends on where the decay happens. It is possible that
such a late decay of dark photons will be treated as missing energy.
If one of the decay products from the heavy resonance is identified
as a photon, while the other one is registered as MET, the
monophoton search can be used to constrain the parameter space. We
will present this constraint in Sec.~\ref{sec:simulation}.

\begin{figure*}
\center
\includegraphics[width=8.5cm]{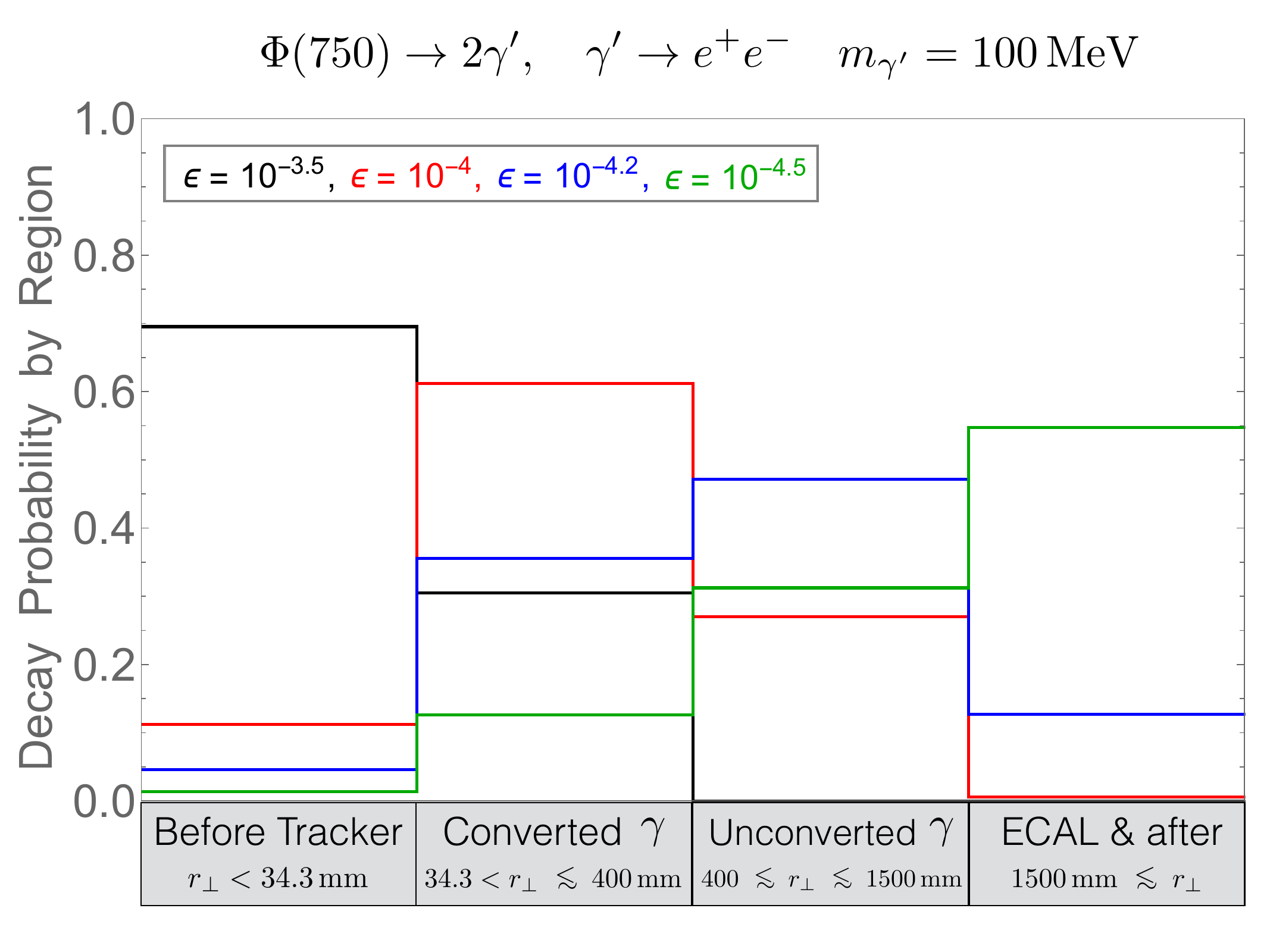}\qquad\includegraphics[width=8.7cm]{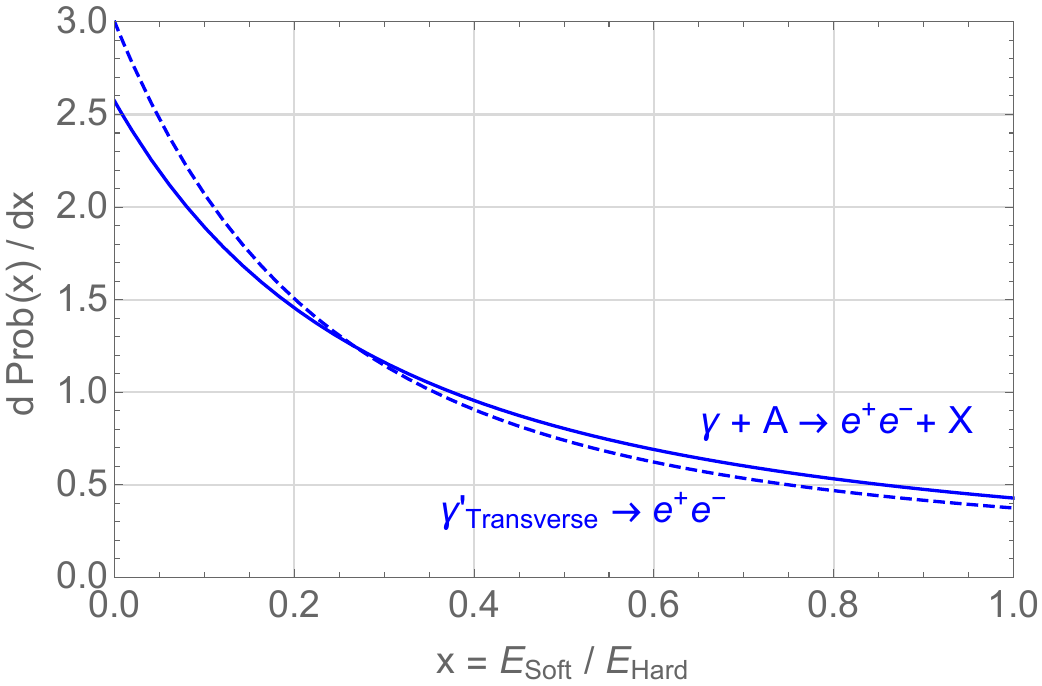}
\caption{Left: Probability of one dark photon decay by regions,
where $r_{\bot}$ is the transverse distance of the decay
vertex from the beam pipe. We take the detailed
positions of the 3rd-to-the-last layer of SCT from
\cite{Aad:2008zzm} in our analysis. The probability for each region
is $\eta$ dependent, and we integrate the result from the central
region to $|\eta|=2.3$ in our estimation based on our simulation.
Right: Comparison of energy ratios between the hard and soft lepton
from the the SM photon conversion and the dark photon decay. The
plot is made by assuming $m_{\gamma'}=100$ MeV, and the decay into
$e^+e^-$ is from the transverse mode of $\gamma'$. When
$m_{\gamma'}\ll E_{\gamma'}$, the distribution is insensitive to the
size of $m_{\gamma'}$.}\label{fig:probregion}
\end{figure*}

In our discussion we focus on the model with
$2m_e<m_\gamma'<2m_\mu$, where a dark photon preferentially decays
into $e^+e^-$. This decay is induced by the
kinetic mixing between the SM photon and an extra $U(1)'$ gauge
boson, i.e. $\mathscr{L}\supset(\epsilon/2) F_{\mu\nu}F'^{\mu\nu}$.
The dark photon, defined in the mass basis after canonical
normalization, has a proper decay length
\begin{eqnarray}
c\tau_{\gamma'\to e^+e^-}&\simeq&\left(\frac{\epsilon^2 \alpha_{\text{EM}}\,m_{\gamma'}}{3}\right)^{-1}
\\&=&8\times 10^{-3}\,\text{cm}\left(\frac{10^{-4}}{\epsilon}\right)^{2}\left(\frac{100\,\text{MeV}}{m_{\gamma'}}\right).\nonumber
\end{eqnarray}
To illustrate the idea with some concrete results, we take the $750$
GeV resonance \cite{ATLAS-CONF-2015-081, CMS-PAS-EXO-15-004,
ATLAS-CONF-2016-059, Khachatryan:2016hje}, which has drawn a lot of
attentions previously, as an example in the collider study. However,
we emphasize that the discussions in this work can be applied for
many photon-related searches at the LHC. In
Fig.~\ref{fig:probregion} (left), we present our simulation results
on the probability of having dark photons with different lifetimes
decay in the four regions discussed above. The plot assumes the dark
photon, which carries a mass $m_{\gamma'}\sim100$ MeV, is generated
from the decay of a $750$ GeV scalar resonance produced by the gluon
fusion. If the heavy resonance is produced by $q-\bar q$ scattering,
it may have a different boost along the beam direction due to
different parton distribution functions of $q\bar q$ and gluon. This
can affect the angular distribution of the final states in the lab
frame, and detector effects can cause a subtlety. We study both
cases and find a negligible difference between the two scenarios.

This result takes into account the geometry of ATLAS
detector at different $\eta$-angle. In principle, one can
study whether there are fake photons from the dark photon decays by
calculating the ratio of the numbers of the converted versus
unconverted photons. Depending on the statistics of photon signals,
we can constrain the decay lifetime by requiring this ratio to be
compatible with the one from real photon events. As is shown in
Fig.~\ref{fig:probregion} (left), this can be satisfied when the
kinetic mixing $\epsilon\sim 10^{-4}$, which is also consistent with
current constraints from various dark photon searches. As the example we
take as a benchmark point, i.e. the diphoton resonance search at
around 750 GeV, the current statistics in the high invariant mass region is not good
enough to impose a strong constraint on $\epsilon$ by checking the
converted/unconverted ratio. Thus there is still a generic choice in
our parameter space so that this ratio is compatible with that of a
real photon. However, one expects that this can be reasonably
improved in the future when having a higher statistics.

One may suggest to distinguish converted photon and light dark
photon decay by comparing the energy ratio between $e^+e^-$. However
we show in the right panel of Fig.~1, if the dark photon from the
750 GeV heavy resonance decay is dominantly in transverse
modes \footnote{This is the case if the dark photon couples to the
750 GeV heavy resonance through a triangle-loop induced by heavy
fermions. The longitudinal mode is suppressed by the small Yukawa
coupling between dark Higgs and heavy charged particle running in
the loop.}, the asymmetry of the two leptons from $\gamma'\to
e^+e^-$ is similar to that in the converted photon process
$\gamma+A\to e^+e^-+Z$, where $A$ and $Z$ stand for the
initial and final states of material that assists
the conversion. This is because the electron from a transversely
produced dark photon decay tends to be parallel or anti-parallel to
the dark photon spin, boosting $\gamma'$ into the lab frame makes
one lepton harder than the other one, which generates the asymmetry
\footnote{The angular distribution of a transversely produced dark
photon decay is $\frac{d\text{Prob}(\theta)}{d\cos\theta}\propto
(1+\cos\theta^2)$ in the center-of-mass frame, where $\theta$ is the
open angle between $e^-$ and the spin direction. Here we assume the
resonance mass $\gg m_{\gamma'}\gg m_e$, and the obtained energy
ratio distribution is $\frac{d\sigma}{dx}\propto
(1-\frac{2x}{(1+x)^2})/(1+x)^2$, with $x\equiv
E_{\text{Soft}}/E_{\text{Hard}}$ in the lab frame.}.

\begin{figure}
\includegraphics[width=8.6cm]{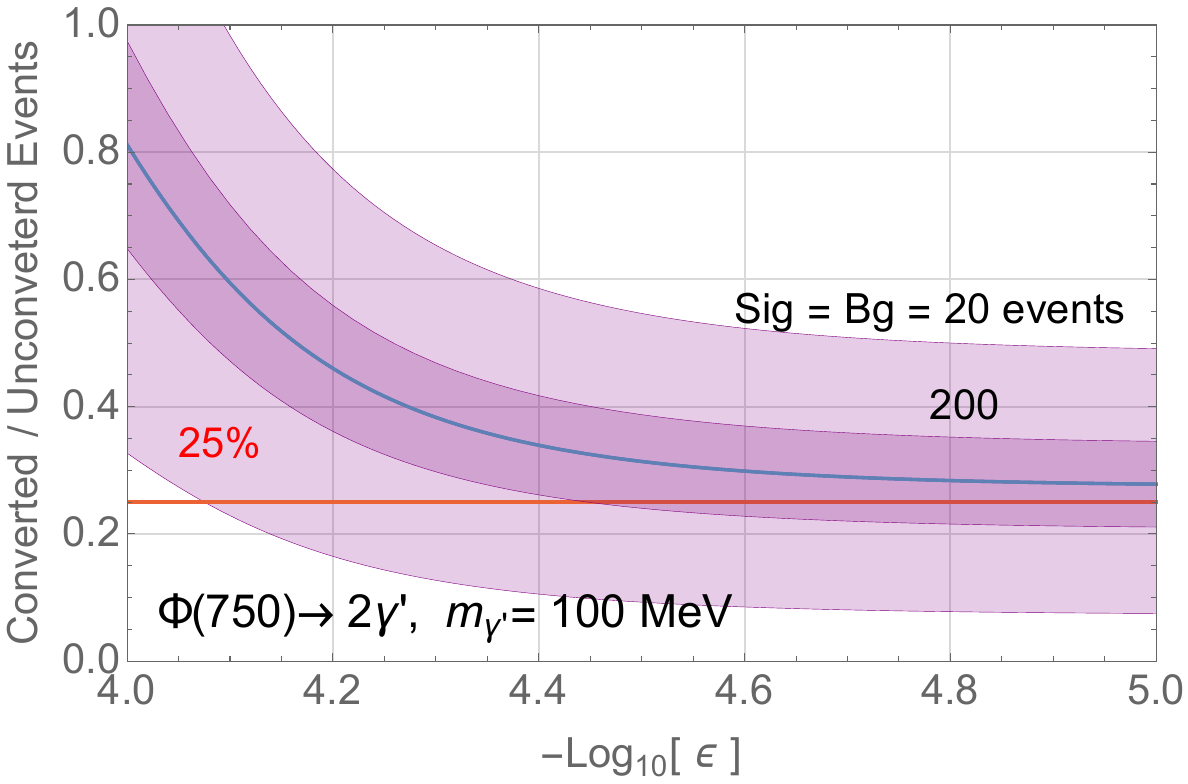}
\caption{Ratio of the fake photon events being identified as
converted and unconverted photons. Blue curve is derived from the
ratio of dark photon decay probability in the converted and
unconverted regions. The light (dark) purple band includes a
$2\sigma$ fluctuation of the photon events as in
Eq.~(\ref{eq:fluctuation}) assuming $20$ ($200$) events of both the
signal and background. According to \cite{conversion}, we take the
ratio of converted vs. unconverted photon events in the SM to be
$25\%$.}\label{fig:conversion}
\end{figure}

Besides distinguishing photons in the inner detector, we may use the showering shape in the ECAL to tell the difference between a single photon versus a highly boosted $\gamma'$ decays into $e^+e^-$. The ``photon jet" objects have been studied for the Higgs search \cite{Ellis:2012zp}, although the focus is on a (pseudo-)scalar particle decaying into two photons. Since the dark photon decay usually deposits most of the photon energy into one of the leptons, the ECAL signal may better resemble a single object. A more detailed study, including the detector response of the electron signal is necessary to identify the dark photon decay.

Another way to distinguish the dark and SM photons is to study the
conversion rate of the photon events. For SM photons, the ratio
$R\equiv N^{\gamma}_{\text{cv}}/N^{\gamma}_{\text{uc}}$ between the
number of converted $N^{\gamma}_{\text{cv}}$ and unconverted
$N^{\gamma}_{\text{uc}}$ events is about $0.25$ when the photon
rapidity $|\eta|<1.5$ \cite{conversion}. In
Fig.~\ref{fig:conversion}, we compare this ratio to the dark photon
events with the event number being equal to the total number of
signals $N^{\gamma'}_{\text{sg}}$ times the probability
$P_{\text{cv}}$ or $P_{\text{uc}}$ of the dark photon being
identified as a converted or unconverted photon, i.e.
$N^{\gamma'}_{\text{cv}/\text{uc}}=P_{\text{cv}/\text{uc}}N^{\gamma'}_{\text{sg}}$.
The ratio
$R_{\gamma'+\gamma}=(P_{\text{cv}}N^{\gamma'}_{\text{sg}}+N^{\gamma}_{\text{cv}})/(P_{\text{uc}}N^{\gamma'}_{\text{sg}}+N^{\gamma}_{\text{uc}})$
is shown in the blue curve in Fig.~\ref{fig:conversion}, which
represents the converted versus unconverted photon ratio without the
presence of other SM background. When kinetic mixing parameter satisfies
$\epsilon\lsim 10^{-4.6}$, the photon decay length is comparable to
the size of the detector, the asymptotic value of the ratio for an even smaller $\epsilon$ is determined by the 
geometry of the detector. To include fluctuation of
the signal and background, we show a colored band of the event ratio
with the upper and lower boundaries, which corresponds to a
$1.64\sigma$ uncertainty that is comparable to the $90\%$ CL
deviation from the expected ratio
\begin{equation}\label{eq:fluctuation}
R^{\pm}_{\gamma'+\gamma}\equiv\frac{N^{\gamma'+\gamma}_{\text{cv}}}{N^{\gamma'+\gamma}_{\text{uc}}}\left(1\pm1.64\sqrt{\left(N_{\text{cv}}^{\gamma'+\gamma}\right)^{-1}+\left(N_{\text{uc}}^{\gamma'+\gamma}\right)^{-1}}\right).
\end{equation}
Here
$N^{\gamma'+\gamma}_{\text{cv}}=P_{\text{cv}}N^{\gamma'}_{\text{sg}}+0.2N^{\gamma}_{\text{bg}}$,
and
$N^{\gamma'+\gamma}_{\text{uc}}=P_{\text{uc}}N^{\gamma'}_{\text{sg}}+0.8N^{\gamma}_{\text{bg}}$.
To illustrate the idea, we assume the production of fake photon
events is of a fb level, and the number of fake photon events is
comparable to the SM photon background to play an important role. In
Fig.~\ref{fig:conversion}, the light purple region assumes
$N^{\gamma'}_{\text{sg}}=N^{\gamma}_{\text{bg}}=20$, which gives an
idea of the $R_{\gamma'+\gamma}$ deviation at the early 13 TeV run
at the LHC. In this case, checking the ratio does not distinguish
the two different photons when $\epsilon\lsim 10^{-4}$. A stronger
result can be obtained when
$N^{\gamma'}_{\text{sg}}=N^{\gamma}_{\text{bg}}=200$ with
$\mathcal{O}(100)$ fb of data.

\begin{figure*}
\center
\includegraphics[width=8.5cm]{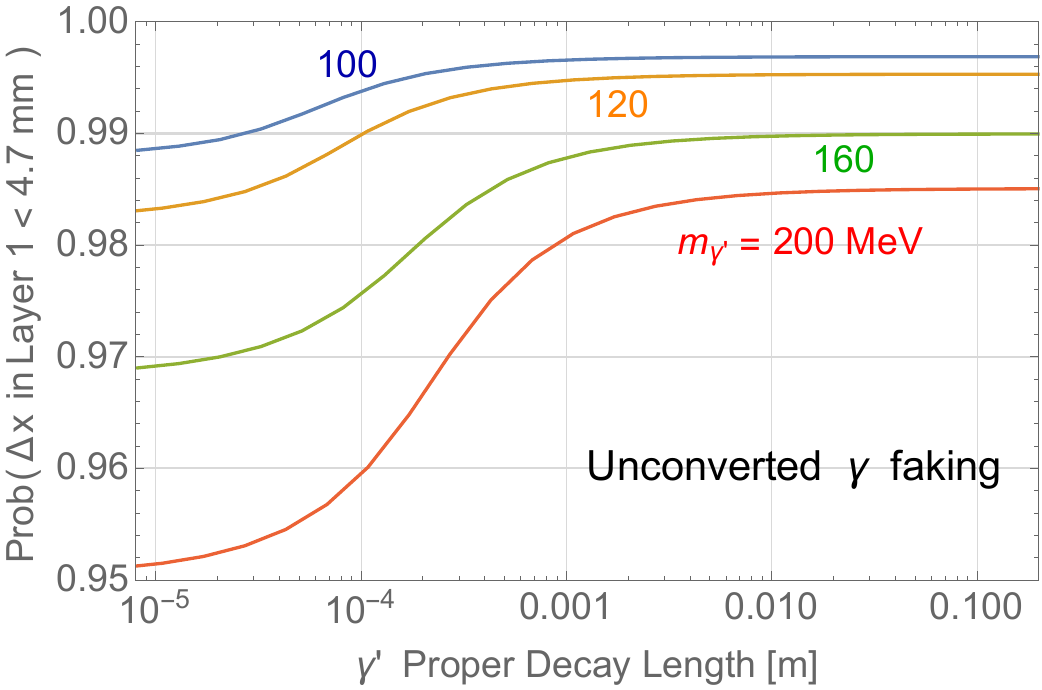}\qquad\includegraphics[width=8.6cm]{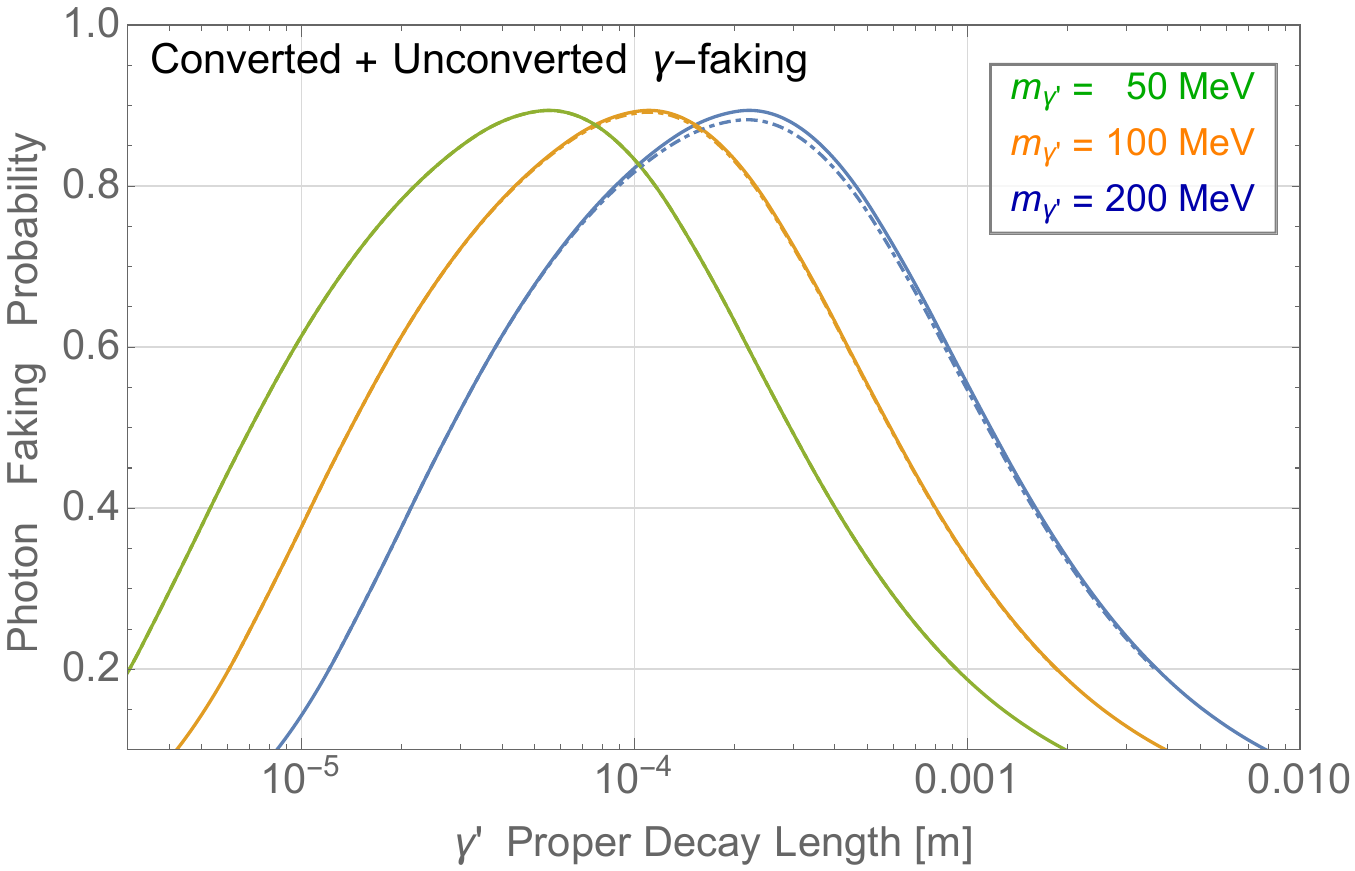}
\caption{Left: Probability of a dark photon produced from
$\Phi(750)\to 2\gamma'$ decaying into $e^+e^-$ between radial
distance $[371,\,1500]$ mm (from the second layer of SCT to ECAL
with $\eta=0$) that has its two hits in the $\eta$-direction of the
strip cells (defined as $\Delta x$) to be within the average
thickness of the cell ($<4.7$ mm). Here we assume the $\gamma'$
moves in the radial ($\eta=0$) direction. Right: Probability of one
dark photon fakes a real photon. The result is derived according to
the decay probability with (solid curve) and without (dash-dotted
curve) the upper bound on the angular separation in the ECAL for the
fake normal photon signals. See Sec.~\ref{sec:detector} for more
discussions. Here we estimate the acceptance of this cut by assuming
dark photons to be produced in the transverse mode. After boosting
the open angle to the lab-frame, we calculate the acceptance as a
function of the decay position.}\label{fig:openangle}
\end{figure*}

In the left panel of Fig.~\ref{fig:openangle}, we show the
efficiency of a dark photon decay in the unconverted photon region
that passes the open angle cut, assuming $\eta(\gamma')=0$ for
simplicity. The cut requires the $e^+e^-$ pair that fakes a normal
photon to hit the first layer of the ATLAS ECAL within a
separation\footnote{This is the average width of a strip cell. The
precise number is angular dependent due to detector details, and we
only take one fixed number in our analysis to simplify the
calculation. The result from this estimation should only be
different from the real answer by a factor of O(1).} $\Delta x< 4.7$
mm in the $\eta$-direction, so the signal can be identified as a
single photon. This is a rather conservative assumption comparing to
the estimation using boosted pions from the Higgs decay measurement
\cite{Knapen:2015dap}. This criteria is easier to be
satisfied for a $\gamma'$ with a larger boost (smaller
$m_{\gamma'}$) or a longer decay length. This
is simply because that, for a fixed value of $\Delta x$, a larger
open angle between $e+/e-$ is allowed if the decay happens closer to
the ECAL.

In the right panel of Fig.~\ref{fig:openangle}, we show the
probability of a dark photon decay being identified as a photon in
the ATLAS detector. The estimation is based on the probability of
dark photon decaying in the converted and unconverted photon regions
(solid curves). The dash-dotted curves in Fig.~\ref{fig:openangle}
contain the angular cut discussed above. As we can see in the plot,
the difference between the probability with (solid curve) and
without (dot-dashed curve) the angular cut is not significant. This
is because an O(100) MeV dark photon is highly boosted from the
decay of a 750 GeV resonance.

\section{Models}\label{sec:models}

As illustrated in sections above, we show how a dark photon can fake
SM photon. In this section, we establish how dark photon can affect
our understanding to underline physics from model building point of
view. Since SM photon and dark photon is not trivial to be
distinguished with current photon identification strategy, we
emphasize that wrong conclusions may be drawn without this subtlety
in mind. To make the physics as clear as possible, we consider the
simplest scenario where a heavy resonance is produced at the LHC and
further decays to two dark photons or one dark photon and one SM
photon. With the current photon ID, such scenario can show up as a
diphoton resonance.  To be noted, such benchmark scenario was
originally motivated by the 750 GeV diphoton excess announced by
ATLAS and CMS at the end of 2015. Although this excess has gone away
in the new set of data, our work remains valid and meaningful since
we are focused on the subtleties on object identification at the LHC
and our study can be generically applied to any scenarios with
photon involved.

First, a heavy resonance decaying to two photons can be introduced
by integrating out charged particles which couple to the heavy
resonance. Since the electromagnetic coupling constant has been
fixed, the signal rate is directly related to the mass of the
charged particles as well as their coupling to the heavy resonance.
Assuming the number of species and the electric charge of charged
particles are not too large, one can derive an upper limit to the
signal rate for such diphoton channel by requiring perturbativity.
On the other hand, if one or both photons are actually dark photons,
the upper bound can be largely relaxed due to undetermined dark
photon coupling constant.

Further, SM photon and Z boson are mixtures of $SU(2)_W$ and
$U(1)_Y$ gauge bosons. If a heavy particle can decay to SM photon,
then very likely, there are channels with Z boson in final states.
Although the detailed numbers on relative decay branching ratios
depend on the representations of particles running in the loop, one
generically expects them to be O(1). However, if the heavy resonance
only decays to two dark photons which fake the photon at the LHC,
then the decay channels with Z boson involved do not necessarily
exist.

It is well known that a heavy vector boson cannot decay to two
on-shell photons, thanks to Landau-Yang theorem. If a heavy
resonance is found in diphoton channel, one naively eliminates the
possibility that such heavy resonance being a spin-1 particle. On
the other hand, a heavy vector boson can decay to a SM photon and a
dark photon, since the final states are not identical particles.
Without understanding the fact that a dark photon can easily fake a
SM photon at the LHC, one can simply miss this possibility and lead
to confusions.

In the following discussions in this section, we discuss more
details of this benchmark model. As stated previously, our paper was
originally motivated by the 750 GeV diphoton excess, we choose our
benchmark point where the resonance is 750 GeV and we assume the
diphoton channel signal rate to be 5 fb at 13 TeV. However, our
conclusion and the physics we discuss do not rely on this excess at
all.


\subsection{Spin-0 resonance}\label{sec:Spin0}
Depending on whether the scalar resonance $\Phi$ is a real
or a pseudo scalar, similar to dilaton or axion Lagrangian, it can
decay into dark photons through an operator,
\begin{eqnarray}
 \label{diphotondecay}
\mathscr{L}_{\gamma\gamma}\supset \frac{g_d^2\,y_{\Phi}}{16\pi^2
M'_1}\Phi F^{'\mu\nu}F_{\mu\nu}^{'} \ \textrm{or} \ \ \
\frac{g_d^2\,y_{\Phi}}{16\pi^2 M'_2}\Phi\ F^{'\mu\nu}\tilde
F'_{\mu\nu}.
\end{eqnarray}
Such operators can be introduced by integrating out particles
charged under a dark U$(1)'$, where $g_d$ is the U$(1)'$ gauge
coupling. $\Phi$ can either be a real or a pseudo scalar. It is also
possible that one of the decay products is a SM photon, so the
possible decay channels are
\begin{equation}\label{eq:Phitogamma}
\Phi(750)\to\gamma'\gamma',\qquad\Phi(750)\to\gamma\gamma'.
\end{equation}
The latter decay, however, may require the existence of a light SM
charged mediator. Actually, in this case, there should be a $\gamma
\gamma$ mode as well. This can be parametrically suppressed by
choosing $e/g_d < 1$. But then the $\gamma' \gamma $ mode is also
suppressed relative to the $\gamma'  \gamma'$ mode.

A scalar resonance can be produced through the gluon fusion, $q-\bar
q$ scattering or electroweak vector boson fusion (VBF). For the VBF
process, the signal usually come with forward jets and is relatively
easy to be tagged on. Further, if the scalar can directly couple to
W or Z bosons at tree level, it has to be a condensed field that
charged under the SM electroweak gauge groups. Such a scenario can
be highly constrained, and we do not consider the VBF production in
the rest of the discussion.

If the heavy resonance couples to gluons through loop diagrams, its
effective coupling can be characterized as
\begin{eqnarray}
 \label{ggproduction}
\mathscr{L}_{gg}\supset \frac{g_3^2\,y_{\Phi}}{16\pi^2 M_1}\Phi\
\textrm{Tr}[G^{\mu\nu}G_{\mu\nu}]\nonumber\\     \textrm{or}\ \
  \frac{g_3^2\,y_{\Phi}}{16\pi^2 M_2}\Phi\
\textrm{Tr}[G^{\mu\nu}\tilde G_{\mu\nu}],
\end{eqnarray}
depending on whether $\Phi$ is a real or pseudo-real scalar boson.
We will assume the diphoton rate to be $5$ fb at 13 TeV. To obtain a
$20$ fb $\Phi(750)$ production with Br$(\Phi\to2\gamma')=0.25$, we
need a SM colored mediator carrying a mass $ M_{1,2}\simeq 3$ TeV,
and a U$(1)'$ charged mediator with mass $ M'_{1,2}\simeq
N_F\,g^2_d\times1.5$ TeV. Here $(N_F,\,g_d)$ are the flavor and
U(1)$^{'}$ coupling of the mediator.

The $\Phi$ can also couple to $q-\bar q$, e.g. by sharing the same
$SU(2)_L$ and U(1)$_Y$ charges as Higgs boson. If $\Phi$ is inert,
so it does not condense or mix with SM Higgs, the Higgs measurement
constraints can be avoided. The decay into dark photons is also not
suppressed by the tree-level process into $W/Z$. In four component
notation after electroweak symmetry breaking, the coupling can be
written as
\begin{eqnarray}
 \label{qqproduction}
\mathscr{L}_{qq}\supset y_{\Phi}\Phi\,\bar q q\  \ \ \ \
\textrm{or}\ \ \ \ \ i y_{\Phi}\Phi\,\bar q\gamma_5 q.
\end{eqnarray}
However, one has to tune the flavor structure in order to obtain a
large enough production while avoid flavor constraints. If one
assumes a flavor diagonal Yukawa couplings with a universal coupling
for all light quarks, having a $20$ fb production and
Br$(\Phi\to2\gamma')=0.25$ requires $M'_{1,2}\simeq N_F\,g^2_d\times
0.7$ TeV.

\subsection{Spin-1 resonance}
If we allow dark photon in the final states, it is also possible for
the heavy resonance being a vector boson, i.e. $V^\mu_H$ as a gauge
boson of a condensed U$(1)_H$. Landau-Yang theorem forbids the
massive vector decaying to two photons. It is possible that the
heavy vector decays to two dark photons because they are not
massless \cite{Toro:2012sv,Chala:2015cev}. However, the decay rate are expected to be highly
suppressed due to the small mass of dark photon. On the other hand,
it is possible that the heavy vector decays to one SM photon and one
dark photon,
\begin{equation}\label{eq:Vtogamma}
V(750)\to \gamma\gamma'.
\end{equation}
In this case, the two particles in the final states are not
identical, thus no more suppression is applied to this decay
channel. The decay of the heavy vector boson can be characterized by
effective operators, for example,
\begin{eqnarray}
 \label{Vectordecay}
\mathscr{L}_{V_H}\supset
\frac{e\,g_d\,g_H}{16\pi^2\tilde{M}_{1}^2}F_{V_H}^{\mu\nu}F_{\nu}^\alpha
F'_{\alpha\mu}
\end{eqnarray}
$g_d$ and $g_H$ are $\gamma'$ and $V_H$ gauge couplings of the heavy
mediator running in the loop. Having a SM photon in the final state
requires the existence of SM charged mediators. Since $V_H$ is a
color singlet, it cannot be produced through gluon fusion due to
Landau-Yang theorem. For a $q-\bar q$ production of $V_H$, the gauge
coupling as large as $\sim 10^{-2}$ can give an $\mathcal{O}(10)$ fb
production of $V_H$ at 13 TeV. In order to obtain a sizable
branching ratio, we need the mass of SM charged mediator to be
$\tilde{M}\sim \sqrt{g_d\,g_HN_F}\times220$ GeV, where $N_F$ is the
number of flavors running in the loop.

Besides the loop-induced decay into (dark) photons, if the dark Higgs
which gives the mass to dark photon is also charged under U$(1)_H$,
$V_H$ can decay into the dark Higgs and the longitudinal mode of
dark photon. Generically, the dark Higgs has a mass close to dark
photon, thus it is kinematically allowed for this dark Higgs to
decay into two dark photons, and this decay can happen at detector
scale. When the decay of the dark Higgs happens inside the region of
the unconverted photon identification, i.e. Region III from
Sec.~\ref{sec:identification}, there is a chance that such process
can fake a SM photon. Since the typical separation between the four
leptons is $\Delta R\sim m_{\phi_d}/2 m_{V_H}\sim 10^{-4}$ for
$m_{\phi_d}=\mathcal{O}(100)$ MeV, the four leptons from the decay
can be within the angular resolution of the ECAL. The benefit of
this scenario is that the heavy vector resonance decays through a
marginal operator and no light charged particles at O(100) GeV needs
to be introduced to UV complete an irrelevant operator. However, the
detector study of such scenario is more complicated, and we leave
the detailed study of the reconstruction efficiency for future work.

\subsection{Resonance as a heavy bound state}
As we have seen in the previous discussion. If the heavy resonance
decays through loop-induced irrelevant operators, a low suppression
scale is usually needed in order to achieve sizable branching ratio
to signal channel. Another interesting possibility is that the heavy resonance is actually a bound state formed by a pair of
new particles. If the new particle has a strong coupling to light
dark photon, there is a high probability that the bound state is
produced instead of two free particles \footnote{See
\cite{Tsai:2015ugz} for more discussions on the bound state
production with self interacting dark matter (SIDM) coupled through
dark photons and \cite{Bi:2016gca} for the discussion of asymmetric
dark matter. Also see \cite{Han:2016aa} for having the $750$ GeV
resonance as a bound state of heavy colored fermions.}. The bound
state can be either a scalar or a vector. If it is a scalar, the
annihilation decay is dominantly by a pair of the force mediator,
i.e. dark photon. The decay width of the bound state can be
estimated as
\begin{eqnarray}
 \label{BoundDecay}
\Gamma\sim \frac{\alpha_{\gamma'}^5}{4}M_B
\end{eqnarray}
where $\alpha_{\gamma'}$ is the dark photon coupling constant and
$M_B$ is the bound state mass. Three powers of
$\alpha_{\gamma'}$ in Eq. (\ref{BoundDecay})
comes from the wavefunction suppression in the bound state. It
appears in all annihilation decay channels. Thus dark photon pair
decay channel is preferred for scalar resonance due to strong
coupling of the dark photon. On the other hand, if the resonance is
a vector state, the $\gamma+\gamma'$ final state is more preferred.

When the heavy resonance is a vector field, one interesting
possibility is that the decay $V(750)\to3\gamma'$ is comparable to
the $\gamma\gamma'$ channel. Although the Landau-Yang theorem
forbids the vector to decay into two SM photons, and the two dark
photon channel is highly suppressed by a power law of $m_{\gamma'}$
over the resonance mass, there is nothing preventing the heavy
resonance from decaying into three dark photons (see also
\cite{Chala:2015cev}). When the dark photon has a much stronger
coupling than that of normal photon, the $3\gamma'$ channel may have
a similar decay branching ratio to $\gamma\gamma'$, and the search
of three photon resonance can be important. For example, as we
discussed previously, the heavy resonance may be a vector bound
state formed by the dark photon exchange. In this case, dark photon
needs to have a large coupling, such as $\alpha_{\gamma'}\sim 0.3$,
in order to have a sizable probability to form a bound state at the
LHC. If both decay processes are allowed, the decay branching ratios
into $3\gamma'$ and $\gamma\gamma'$ are compatible,
\begin{eqnarray}
 \label{Compare}
\frac{\Gamma_{V_H\to 3\gamma'}}{\Gamma_{V_H\to
\gamma+\gamma'}}\sim\frac{\alpha_{\gamma'}}{4\pi}\frac{\alpha_{\gamma'}}{\alpha}\simeq
\left(\frac{\alpha_{\gamma'}}{0.3}\right)^2,
\end{eqnarray}
and the dark photons in the final states will further decay into
collimated $e^+e^-$ that can fake the SM photons. Searching for a
resonance in 3-photon final states may provide an interesting probe
to this possibility. On the other hand, the boost factor of dark
photon from 3-body decay is smaller than that from 2-body decay.
Thus the dark photon will have a shorter decay length in the lab
frame. This may change the efficiency of being identified as a
photon, and a detailed simulation is necessary to take into account
the detector effects.

\section{Simulation Results}\label{sec:simulation}
Let us first focus on the scenario where the resonance is a scalar
boson. There is a generic choice of parameter space, i.e.
$(m_{\gamma'},\,\epsilon)$, for a dark photon to fake a SM photon
with a probability $\gsim 80\%$. To estimate the probability, we
first simulate a parton-level process $gg\to\Phi(750)\to 2\gamma'$
using MadGraph5.2 \cite{Alwall:2014hca} and the effective coupling
in Eq.~(\ref{diphotondecay}) coded using FeynRules 2.3
\cite{Alloul:2013bka}. We carry out our analysis at parton level and
we do not expect the conclusion to change much after detailed
detector simulations are included. We apply kinematic cuts
$p_{T}(\gamma')>300$ GeV ($0.4\,m_{\gamma\gamma}$ taken in
\cite{ATLAS-CONF-2015-081}), $|\eta(\gamma')|<2.37$ in event
selections. For each dark photon, we take its
$(E_{\gamma'},\,\eta_{\gamma'})$ and calculate the corresponding
decay probability in each photon identification region, as described
in Sec.~\ref{sec:identification}. We carry out similar analysis for
$q\bar q\to V(750)\to \gamma\gamma'$ as well. The energy
calibrations on converted and unconverted photons are both small,
i.e. $|E/E_{\text{true}}-1|\lsim 0.5\%$ from the ATLAS study
\cite{Aad:2014nim}, and we do not expect such calibrations to cause
additional subtleties when determining the dark photon energy.

\begin{figure}
\center
\includegraphics[width=8.5cm]{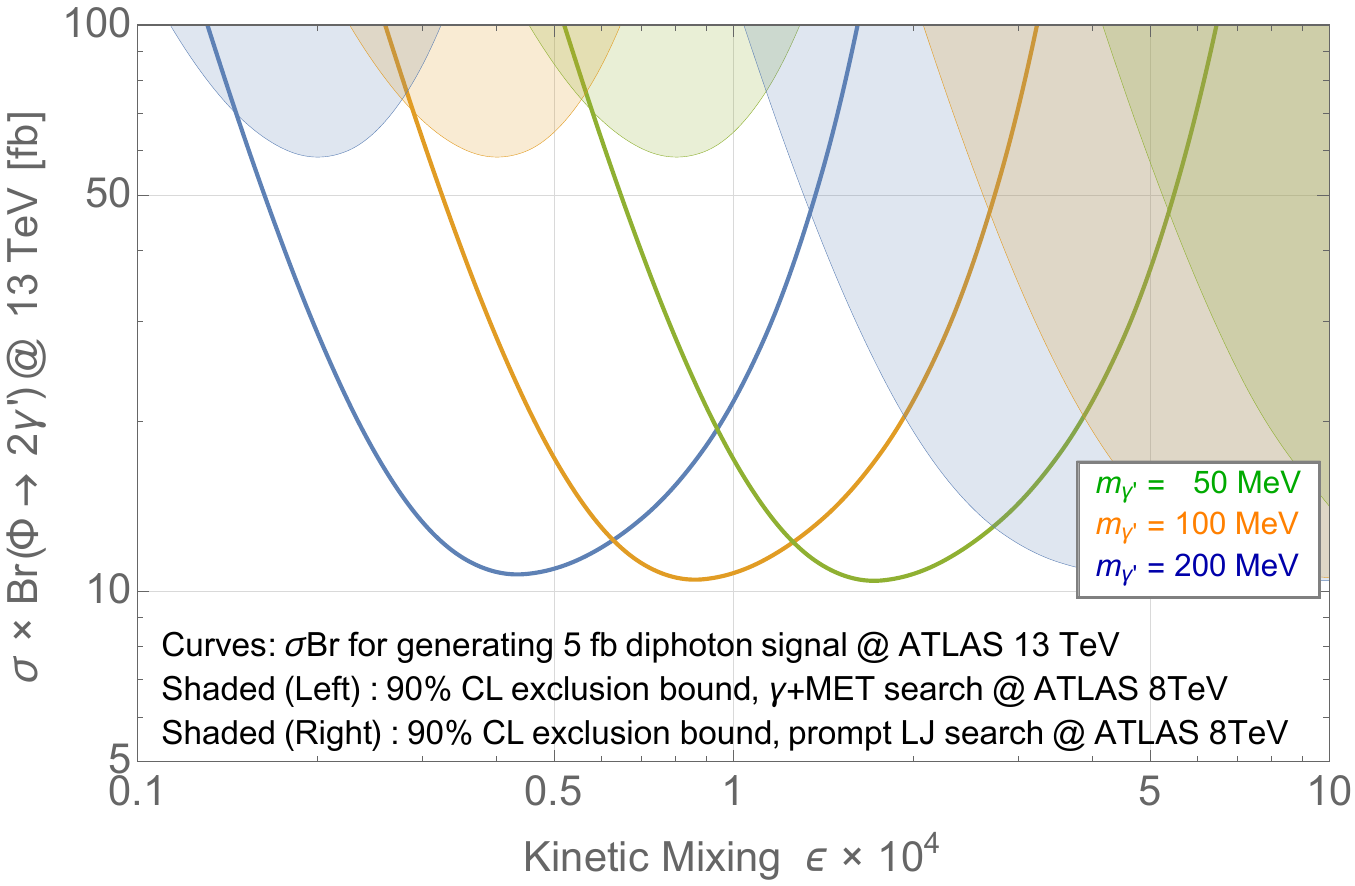}
\caption{The required cross section times branching ratio of the
scalar resonance $\Phi(750)\to2\gamma'$ that generates a $5$ fb
diphoton signal at the $13$ TeV ATLAS search, assuming
Br$(\gamma'\to e^+e^-)=1$. Different color schemes represent various
dark photon masses $m_{\gamma'}$, and the shaded regions give the
$90\%$ CL exclusion bounds from the $8$ TeV ATLAS search of a
photon$+$MET  \cite{Aad:2014tda} (exclusion regions on the left) and
prompt lepton jets \cite{Aad:2015sms} (exclusion regions on the
right). Here we carry out the analysis conservatively by assuming
the highly collimated $e^+e^-$ from the light $\gamma'$ decays
cannot be reconstructed if the decay happens after the first layer
of the ECAL ($\lsim 1590$ mm). Since the MET in \cite{Aad:2014tda} is
defined as the imbalance of the total $p_T$ of the constructed
objects, we treat such displaced decay as missing energy. When the
decay happens before Pixel ($\lsim 34$ mm), the highly collimated
lepton pairs are identified as lepton jets. The rescaling of the
resonance production between $13/8$ TeV is based on the
$gg\to\Phi(750)$.}\label{fig:rate_constraint}
\end{figure}

In Fig.~\ref{fig:rate_constraint}, we study $gg\to\Phi(750)\to
2\gamma'$ process and show the required resonance production with
$5$ fb diphoton signal at $13$ TeV in ATLAS. As expected, changing
kinetic mixing affects the event selection acceptance, and the total
production cross section needs to be modified accordingly in order
to give $5$ fb signal rate in diphoton channel. Such a signal rate can
be easily obtained by a generic choice of the parameter space.

If the dark photon decays after the ECAL, there is a good chance for the
signals to be either un-constructible (gives missing energy) or
identified as displaced lepton-jets \cite{Aad:2014yea}. As we
discuss in Sec.~\ref{sec:detector}, searches of displaced lepton-jets
does not yet provide relevant constraints. In the monophoton search
\cite{Aad:2014tda}, the MET is defined as the imbalance of the total
$p_T$ of constructed objects. If we take the conservative assumption
by identifying the decays after the first layer of the ECAL to be
contributions to MET, monophoton search can be applied to constrain
the parameter space \cite{Khachatryan:2015qba,Aad:2014tda}. In
Fig.~\ref{fig:rate_constraint}, the shaded region on the top left is
given by the $90\%$ CL exclusion from monophoton seach at the $8$
TeV search in \cite{Aad:2014tda}. Here we adapt cuts used in the
ATLAS search, i.e. requiring one of the dark photons to have
$p_T>150$ GeV, $|\eta|<1.37$ and decays inside the photon
identification region, while the other photon decays after the first
layer of the ECAL. The bounds are barely relevant for the parameter
space which gives 5 fb signal rate in diphoton channel.

If dark photons decay before the first layer of Pixel ($r\lsim 34$
mm) in the tracker. Such highly collimated $e^+e^-$ from the dark photon
decay can be identified as a lepton-jet. ATLAS carries out a search
on prompt lepton-jet pair \cite{Aad:2015sms}. This search can be
applied to constrain our scenario since it is possible that both
dark photon decay before getting into the tracker, especially when the
mixing $\epsilon$ is large. We show the existing bound on the cross section in
the shaded region at the right side of
Fig.~\ref{fig:rate_constraint}. Lepton-jets are required to have
$p_T>5$ GeV, $|\eta|<2.5$, and we apply a reconstruction factor
$0.6$ for each of the jet which is comparable to the value used in
\cite{Aad:2015sms}. As we see, the bound is only relevant when the
mixing is large and is not useful to constrain most part of the interesting
parameter space.

\begin{figure*}
\center
\includegraphics[width=8cm]{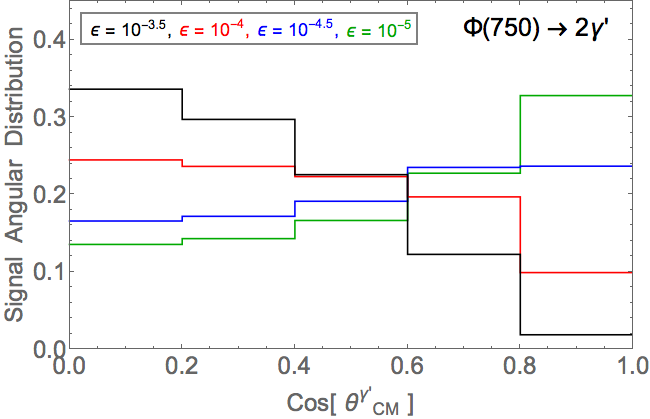}\qquad\includegraphics[width=8cm]{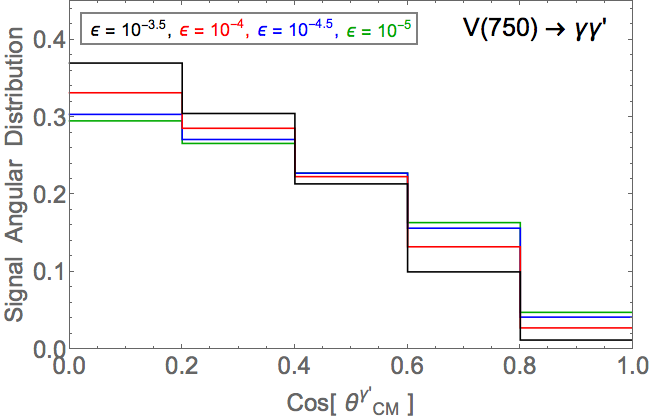}
\caption{Here we take the events which pass a mild diphoton $p_T$
cut, as discussed in context. We show the angular distribution of
the decay products in the rest frame of the heavy resonance, for
scalar (left) and vector (right). Different colors correspond to
various choices of $\epsilon$. We take $m_{\gamma'}=100$ MeV as our
benchmark.}\label{fig:angulartot1}
\end{figure*}

\begin{figure*}
\center
\includegraphics[width=8cm]{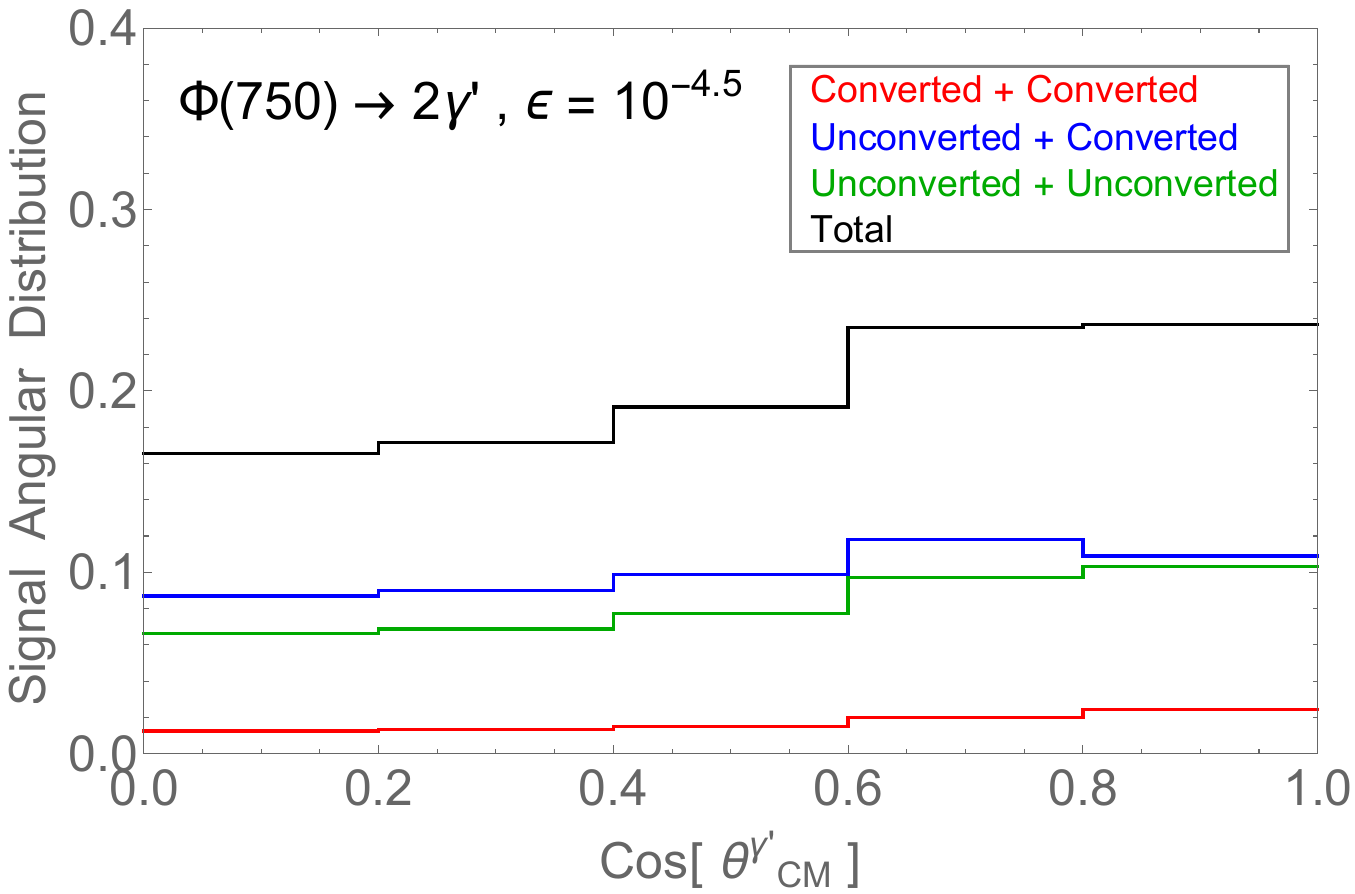}\qquad\includegraphics[width=8cm]{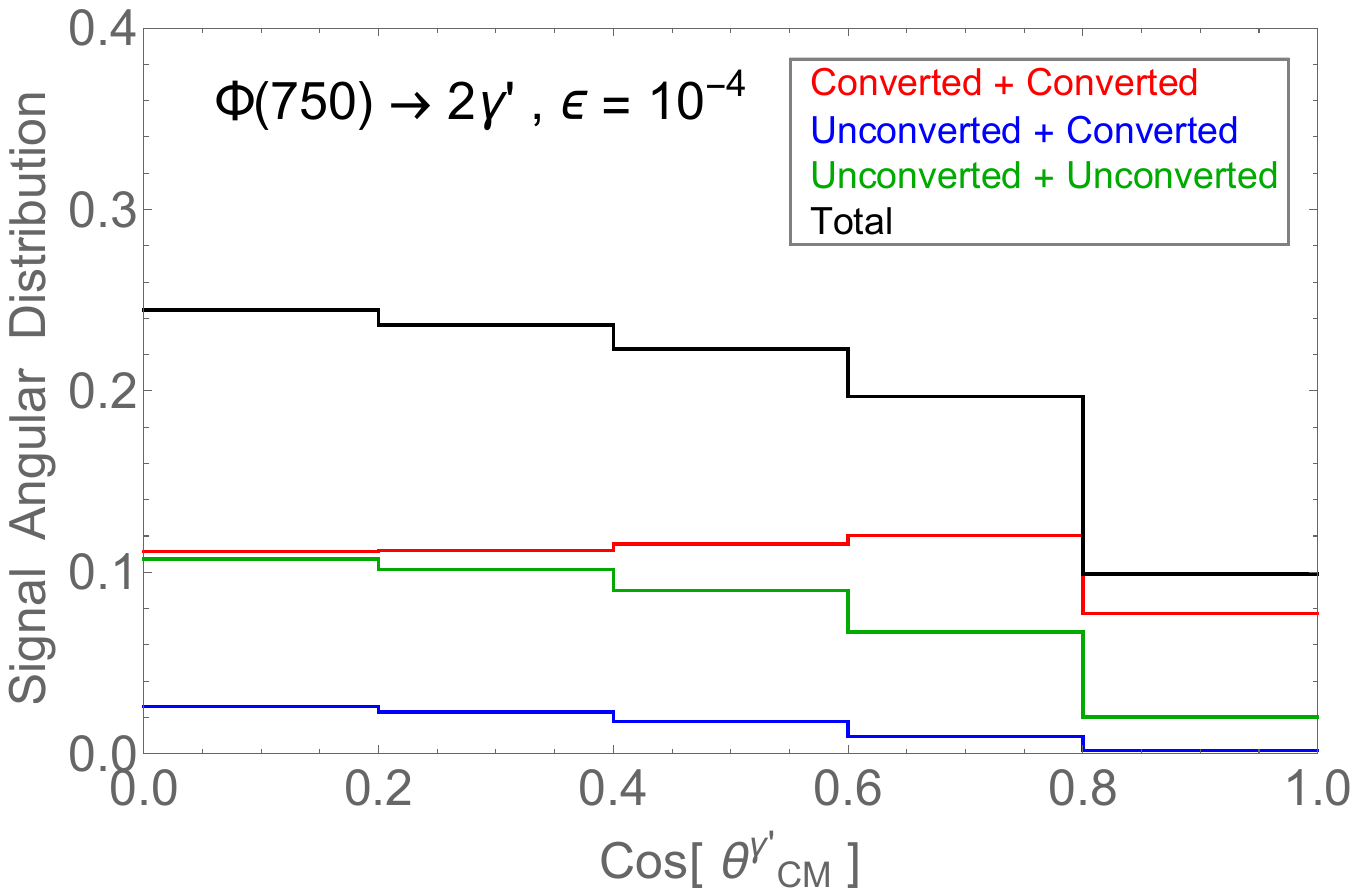}
\\
\includegraphics[width=8cm]{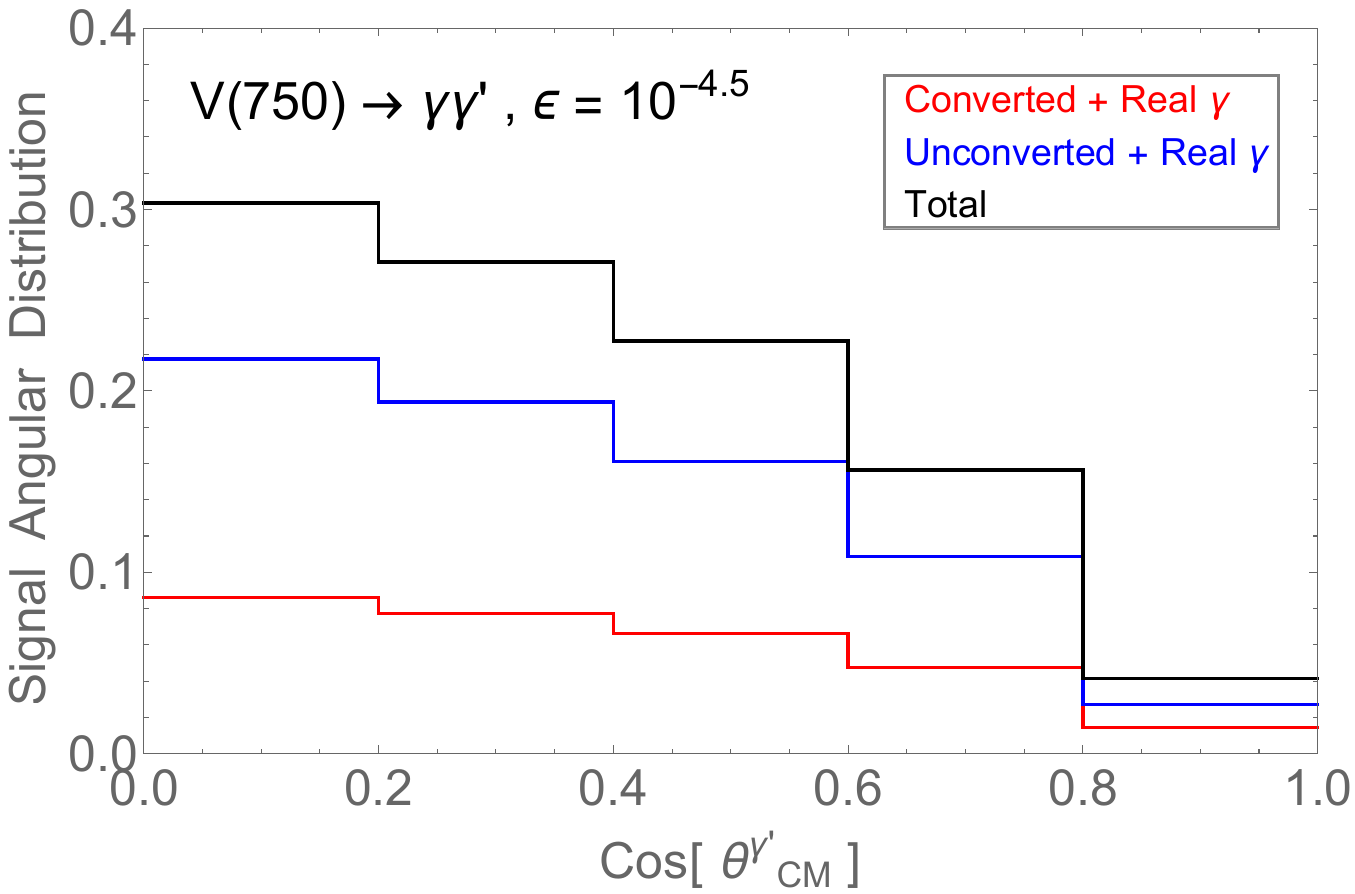}\qquad\includegraphics[width=8cm]{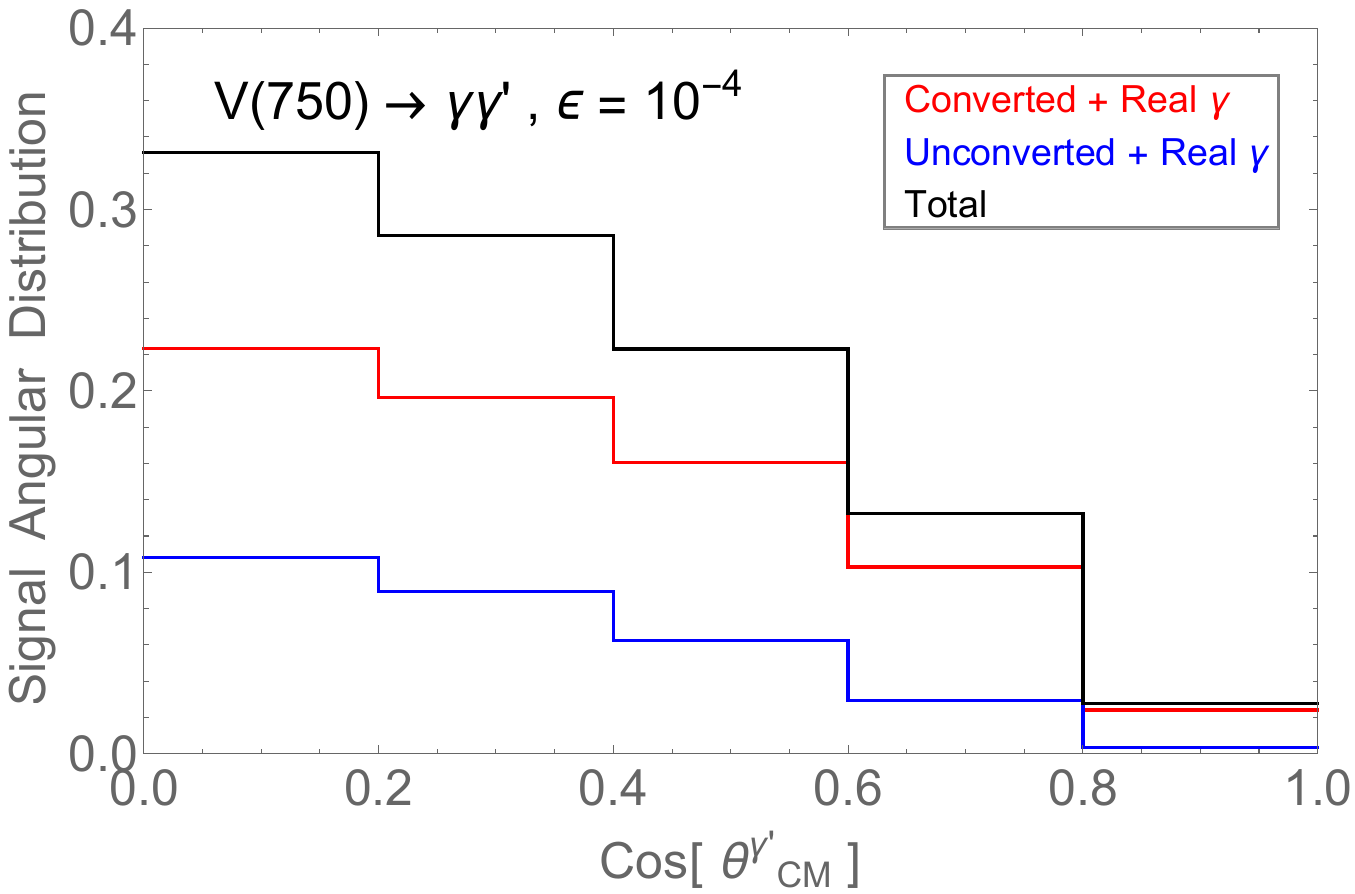}
\caption{The angular distribution of the decay products in the rest
frame of the heavy resonance, assuming $m_{\gamma'}=100$ MeV.
Different colors specify various categories of photon
identification, and the different angular distributions of the
signal may provide a handle to distinguish the real and fake photon
events. When generating the $V\to\gamma\gamma'$
events, we assume the process is dominated by the transverse
component of the massive vectors $V$ and $\gamma'$. Note, the scalar decay is a flat distribution in the
rest frame before taking into consideration of angular dependent
acceptance from the detector. The spin of the vector resonance has a
large probability to point along to the beam direction. Thus the
vector bosons in the final states are more concentrated in central
region in the rest frame, i.e. smaller
$\cos(\theta^{\gamma'}_{\text{CM}})$.}\label{fig:angulartot2}
\end{figure*}

It is interesting to study the angular distribution of the fake
photons. When the decay length in the lab frame is comparable to the
detector size, the geometry of the tracker and ECAL affects the
photon faking probability and introduces detector effects to the
angular distribution of the signals. The current diphoton analysis
in ATLAS imposes hard $p^{\gamma}_T$ cuts in the event selections, which
remove the decays close to the beam direction. However, a wider range of the
angular distribution is necessary for studying the
spin of the heavy resonance, and it is useful to relax the $p^{\gamma}_T$ cut. When studying the
angular distribution in Fig.~\ref{fig:angulartot1} and \ref{fig:angulartot2}, we impose the $p_T$ cut in a milder way by requiring $p^{\gamma}_T>50$ GeV for both photons.

In Fig.~\ref{fig:angulartot1} and \ref{fig:angulartot2}, we take the
events which are registered as diphoton under the relaxed $p_T$
cuts. We boost the events back to the rest frame of the heavy
resonance and study the angular distribution of the decay. For both
the scalar and vector resonances, the signal becomes more forward
when the dark photon has a longer decay length. A larger distance between the interaction point and the ECAL
gives the dark photon more chance to decay in the right region.

The angular distribution is the way to determine the spin of
a particle. In our scenario, the detector effect can easily
introduce a bias to the final state distribution which could cause
confusion or even mistakes when studying the spin of heavy resonance without
bearing the subtlety in mind. Furthermore, as shown in
Fig.~\ref{fig:angulartot2}, angular distribution of various
categories of photon identification can be different. This may
provide a unique way to distinguish a real photon from the displaced
dark photon decay.

\section{Discussion}\label{sec:discussion}
Object identification in high energy colliders is subtle. In this
paper, we study the possibility of faking photon signals at the LHC
using displaced decays of dark photon into $e^+e^-$.  Under a
generic choice of parameters, a dark photons have a large chance to
be identified as a SM photon in detectors. Especially, the decay of
a dark photon cannot only fake a converted but also an unconverted
photon.

When a dark photon decays at different region of the detector, it
will be identified as different object according to the object
reconstruction strategies. This introduces a detector structure
dependence into object identification. More explicitly, given a
fixed decay distance, a dark photon propagating along different
angular direction may decay at different part of the detector, either
before the tracker, within the tracker, within the ECAL or outside
the ECAL. Thus this dark photon decay can be identified as different
objects, such as a lepton jet, a converted photon, an unconverted photon,
or even missing energy. The non-trivial angular dependence in the signal
efficiency that is generated by the detector structure will further
affect the event selection. This can change the result of spin determination of
the heavy resonance. More interestingly, since the event selection is sensitive to
detailed structures of a detector, the probabilities of faking a photon can be
different between ATLAS and CMS. Without unfolding the detector effects,
ATLAS and CMS may observe different number of fake photon events. Comparing
signal cross section or angular distribution between these two
experiments may provide a useful and novel handle to distinguish our
scenario from conventional ones.

A dark photon which decays within the tracker can fake a real
photon. However, that implies a non-negligible probability for the
decay to be before or after the tracker. If it decays before the
tracker, such an object may be identified as a lepton-jet. Further
if the decay happens after the ECAL, it can be identified as
different objects depending on the detailed decay location. If the
decay happens inside the HCAL, it may be identified as a jet with no
electromagnetic (EM) energy. Such exotic objects have already been
covered in some searches \cite{ATLAS-CONF-2014-041,Aad:2014yea}.
Future improvements, such as triggering on one hard photon and one
jet with no EM energy, can be applied in order to increase the
sensitivity. If the decay happens in the muon chamber, another
specialized displace signal search can be imposed to cover the
scenario. Besides to ATLAS and CMS experiments, other dark
photon searches can also provide complimentary constraints to
similar region of dark photon parameters. For instance, a dark
photon with $\mathcal{O}(100)$ MeV mass and $\mathcal{O}(10^{-4})$
mixing can easily fake SM photon in the benchmark model in this
paper. At the meanwhile, the dark photon search at the LHCb
\cite{Ilten:2015hya} can cover similar parameter region. However, we
note that our dark photon production is based on the decay of a
heavy resonance. While in \cite{Ilten:2015hya} and most of other
dark photon searches, the production is only through kinetic mixing.
Depending on the production rate the this heavy resonance, we may or
may not have an earlier coverage on the similar dark photon
parameter region. However, let us emphasize that constraining dark
photon parameter space is not our key focus. The goal of this paper
is to point out the subtleties in particle identification strategies
and study their potential impacts when extracting physics
information in certain searches.

Besides the exotic object search, there could be other
ways to distinguish the dark photon scenario from other possibilities.
In most models where the photons are directly produced from a heavy
resonance decay, other di-boson channels such as $Z\gamma$ and $ZZ$
have non-negligible rates. This is because the particles mediating
the diphoton decay channel also couple to $Z$ boson. However, this
relation does not necessarily hold if the heavy resonance decays
dominantly to two dark photons. The decay branching ratios to
channels with one or two $Z$ bosons can be highly suppressed by
powers of $\epsilon$'s from the kinetic mixing. Comparing other
di-boson decay channels with the diphoton channel then serves a
powerful probe to distinguish models where photons are directly
produced from heavy resonance decay or faked by meta-stable particle
decays.

Although in this paper we focus on
heavy resonance decaying into dark photons, our study really
highlights the possibility of faking photons using dark photon decays.
The lesson one can draw here has a much broader
implication. Dark photons can provide alternative
interpretations of many searches of new physics with photon(s) in
the final state. On the other hand, when searching dark photons, we
may also miss the signal by mis-tagging the displaced dark photon decay
as a regular photon. It is important to explore such a possibilities and develop more sophisticated strategies to distinguish these fake objects.

\section*{Acknowledgments} We are grateful to Alberto Belloni, David Curtin, Michael Mulhearn, Myeonghun Park, and Haichen Wang for useful discussions. Y.T. is supported by the National Science Foundation Grant No. PHY-1315155, and by the Maryland Center
for Fundamental Physics. LTW is supported by DOE grant DE-SC0013642. YZ is supported by DOE grant
DE- SC0007859.
\bibliography{./Diphoton}

\end{document}